\documentclass[aps,prd,12pt,groupedaddress,preprintnumbers,floatfix,nofootinbib]{revtex4}

\usepackage{amsmath}
\usepackage{graphicx}
\usepackage{color}
\usepackage{dcolumn}
\usepackage{bm}

\begin{document}
 \preprint{UCRHEP-T484}
\preprint{DCP-10-01}

\title{Dark Left-Right Gauge Model: SU(2)$_R$ Phenomenology}

\author{
Alfredo Aranda,$^{1,2}$\footnote{Electronic address:fefo@ucol.mx} 
J. Lorenzo D\'iaz-Cruz,$^{2,3}$, Jaime Hern\'andez-S\'anchez,$^{2,4}$ and Ernest Ma$^5$} 

\affiliation{$^1$Facultad de Ciencias - CUICBAS,\\ 
  Universidad de Colima, M\'exico\\
  $^2$Dual C-P Institute of High Energy Physics, M\'exico \\
  $^3$C.A. de Particulas, Campos y Relatividad,
  FCFM-BUAP, Puebla, Pue., Mexico\\
  $^4$ Facultad de Ciencias de la Electr\'onica, BUAP, Avenida San Claudio y 18 Sur, C. P. 72500, Puebla, Pue., M\'exico \\
  $^5$ Department of Physics and Astronomy, University of California,\\
Riverside, California 92521, USA}

\date{\today}

\begin{abstract}
In the recently proposed dark left-right gauge model of particle interactions, 
the left-handed fermion doublet $(\nu,e)_L$ is connected to its right-handed 
counterpart $(n,e)_R$ through a scalar bidoublet, but $\nu_L$ couples 
to $n_R$ only through $\phi_1^0$ which has no vacuum expectation value.  
The usual $R$ parity, i.e. $R = (-)^{3B+L+2j}$, can be defined for this 
nonsupersymmetric model so that both $n$ and $\Phi_1$ are odd together with 
$W_R^\pm$.  The lightest $n$ is thus a viable dark-matter candidate (scotino). 
Here we explore the phenomenology associated with the $SU(2)_R$ gauge group 
of this model, which allows it to appear at the TeV energy scale. The 
exciting possibility of $Z' \to 8$ charged leptons is discussed.
\end{abstract}

\maketitle

\section{Introduction}
The nonsupersymmetric dark 
left-right model (DLRM) proposed recently \cite{klm09} is a variant of a 
supersymmetric left-right extension of the standard model (SM) of particle 
interactions based on $E_6$ and inspired by string theory some 23 years ago
\cite{m87,bhm87}.  It has a number of desirable properties, the chief of 
which is the absence of tree-level flavor-changing neutral currents, thus 
allowing the $SU(2)_R$ breaking scale to be as low as experimentally allowed 
by collider data. This became known in the literature as the alternative 
left-right model (ALRM) \cite{hr89}.  Here we explore further consequences 
of the DLRM, coming from the $SU(2)_R$ sector.

\section{Model}
Consider the gauge group $SU(3)_C \times 
SU(2)_L \times SU(2)_R \times U(1) \times S$, where $S$ is a global symmetry 
such that the breaking of $SU(2)_R \times S$ will leave the combination 
$L = S - T_{3R}$ unbroken.  This allows $L$ to be a generalized lepton 
number which is conserved \cite{klm09} in all interactions except those 
which are responsible for Majorana neutrino masses.  The fermion content 
of the DLRM is given by
\begin{eqnarray}
&& \psi_L = \left(\begin{array}{c}\nu \\ e \end{array}\right)_L  \sim (1,2,1,-1/2;1), ~~~ 
\psi_R = \left(\begin{array}{c}n \\ e\end{array}\right)_R \sim (1,1,2,-1/2;1/2), \\ 
&& Q_L = \left(\begin{array}{c}u \\ d\end{array}\right)_L  \sim (3,2,1,1/6;0), ~~~ d_R \sim (3,1,1,-1/3;0), \\
&& Q_R = \left(\begin{array}{c}u \\ h\end{array}\right)_R  \sim (3,1,2,1/6;1/2), ~~~ h_L \sim (3,1,1,-1/3;1).
\end{eqnarray}
This basic structure was already known many years ago \cite{rr78,bm88} but 
without realizing that $n$ is a scotino, i.e. a dark-matter fermion.
  
The scalar sector of the DLRM consists of one bidoublet and two doublets:
\begin{equation}
\Phi = \begin{pmatrix}\phi_1^0 & \phi_2^+ \cr \phi_1^- & \phi_2^0\end{pmatrix}, ~~~
\Phi_L = \begin{pmatrix}\phi_L^+ \cr \phi_L^0\end{pmatrix}, ~~~ \Phi_R = \begin{pmatrix}\phi_R^+
\cr \phi_R^0\end{pmatrix},
\end{equation}
as well as two triplets for making $\nu$ and $n$ massive separately:
\begin{equation}
\Delta_L = \begin{pmatrix}\Delta_L^+/\sqrt{2} & \Delta_L^{++} \cr \Delta_L^0 &
-\Delta_L^+/\sqrt{2}\end{pmatrix}, ~~~ \Delta_R = \begin{pmatrix}\Delta_R^+/\sqrt{2} &
\Delta_R^{++} \cr \Delta_R^0 & -\Delta_R^+/\sqrt{2}\end{pmatrix}.
\end{equation}
Their assignments under $S$ are listed in Table I.

\begin{table}[htb]
\caption{Scalar content of proposed model.}
\begin{center}
\begin{tabular}{|c|c|c|}
\hline
Scalar & $SU(3)_C \times SU(2)_L \times SU(2)_R \times U(1)$ & $S$ \\
\hline
$\Phi$ & $(1,2,2,0)$ & $1/2$ \\
$\tilde{\Phi} = \sigma_2 \Phi^* \sigma_2$ & $(1,2,2,0)$ & $-1/2$ \\
$\Phi_L$ & $(1,2,1,1/2)$ & $0$ \\
$\Phi_R$ & $(1,1,2,1/2)$ & $-1/2$ \\
$\Delta_L$ & $(1,3,1,1)$ & $-2$ \\
$\Delta_R$ & $(1,1,3,1)$ & $-1$ \\
\hline
\end{tabular}
\end{center}
\end{table}

The Yukawa terms allowed by $S$ are then $\overline{\psi}_L \Phi \psi_R$,
$\overline{Q}_L \tilde{\Phi} Q_R$, $\overline{Q}_L \Phi_L d_R$, $\overline{Q}_R
\Phi_R h_L$, $\psi_L \psi_L \Delta_L$, and $\psi_R \psi_R \Delta_R$, whereas
$\overline{\psi}_L \tilde{\Phi} \psi_R$, $\overline{Q}_L \Phi Q_R$, and
$\overline{h}_L d_R$ are forbidden.  Hence $m_e$, $m_u$ come from $v_2 =
\langle \phi_2^0 \rangle$, $m_d$ comes from $v_3 = \langle \phi_L^0 \rangle$,
$m_h$ comes from $v_4 = \langle \phi_R^0 \rangle$, $m_\nu$ comes from $v_5 =
\langle \Delta_L^0 \rangle$, and $m_n$ comes from $v_6 = \langle \Delta_R^0
\rangle$.  This structure shows clearly that flavor-changing neutral currents
are guaranteed to be absent at tree level~\cite{gw77}.

The generalized lepton number $L = S - T_{3R}$ remains 1 for $\nu$ and $e$, 
and 0 for $u$ and $d$, but the new particle $n$ has $L=0$ and $h$ has $L=1$, 
whereas $W_R^\pm$ has $L=\mp 1$ and $Z'$ has $L=0$, etc.  As neutrinos acquire 
Majorana masses, $L$ is broken to $(-)^L$.  The generalized $R$ parity is 
then defined in the usual way, i.e. $(-)^{3B+L+2j}$.  The known quarks and 
leptons have even $R$, but $n$, $h$, $W_R^\pm$, $\phi_R^\pm$, $\Delta_R^\pm$, 
$\phi_1^\pm$, Re($\phi_1^0$), and Im($\phi_1^0$) have odd $R$.  Hence the
lightest $n$ can be a viable dark-matter candidate if it is also the lightest
among all the particles having odd $R$.  Note that $R$ parity has now been
implemented in a {\it nonsupersymmetric} model.

\section{$SU(2)_R$ Higgs structure}
There exists an experimental bound~\cite{klm09} on $M_{Z'}$ of $850$~GeV from 
Tevatron data~\cite{cdf09}.  As for the recent CDMS-II 
results~\cite{Ahmed:2009zw}, they impose no additional constraint because $n$ 
is Majorana and does not contribute to the s-wave elastic spin-independent 
scattering cross section through $Z'$ exchange in the nonrelativistic limit.  
Assuming thus that $M_{Z'} > 850$ GeV only, we study the $SU(2)_R$ Higgs 
structure of this model and identify those new particles which may be 
relatively light and be observable at the Large Hadron Collider (LHC).  
Consider then the most general Higgs potential consisting of $\Phi_R$ and 
$\Delta_R$:

\begin{eqnarray} \label{potential}
  V_R &=& m_4^2 \Phi_R^\dagger \Phi_R + m_6^2 Tr (\Delta_R^\dagger \Delta_R) + 
  \frac{1}{2} \lambda_1 (\Phi_R^\dagger \Phi_R)^2 + \frac{1}{2} \lambda_2 
  [Tr (\Delta_R^\dagger \Delta_R)]^2 \nonumber \\ &+& \frac{1}{4} \lambda_3 
  Tr (\Delta_R^\dagger \Delta_R - \Delta_R \Delta_R^\dagger)^2 + f_1 (\Phi_R^\dagger 
  \Phi_R) Tr (\Delta_R^\dagger \Delta_R) \nonumber \\ &+& f_2 \Phi_R^\dagger 
  (\Delta_R^\dagger \Delta_R - \Delta_R \Delta_R^\dagger) \Phi_R  + \mu 
  (\Phi_R^\dagger \Delta_R \tilde \Phi_R + \tilde \Phi_R^\dagger \Delta_R^\dagger 
  \Phi_R),
\end{eqnarray}
where
\begin{eqnarray}
\Phi_R^\dagger \Phi_R &=& \phi_R^- \phi_R^+ + \bar \phi_R^0 \phi_R^0, \\ 
Tr (\Delta_R^\dagger \Delta_R) &=& \Delta_R^{--} \Delta_R^{++} + \Delta_R^- 
\Delta_R^+ + \bar \Delta_R^0 \Delta_R^0, \\ 
\Delta_R^\dagger \Delta_R - \Delta_R \Delta_R^\dagger &=&  
\begin{pmatrix}\bar \Delta_R^0 \Delta_R^0 - \Delta_R^{--} \Delta_R^{++} & \sqrt{2} 
(\Delta_R^- \Delta_R^{++} - \bar \Delta_R^0 \Delta_R^+) \cr \sqrt{2} 
(\Delta_R^{--} \Delta_R^+ - \Delta_R^- \Delta_R^0) & - \bar \Delta_R^0 
\Delta_R^0 + \Delta_R^{--} \Delta_R^{++}\end{pmatrix}, \\ 
\tilde \Phi_R^\dagger \Delta_R^\dagger \Phi_R &=& \phi_R^0 \phi_R^0 \bar 
\Delta_R^0 + \sqrt{2} \phi_R^0 \phi_R^+ \Delta_R^- - \phi_R^+ \phi_R^+ 
\Delta_R^{--}.
\end{eqnarray}
Let $\langle \phi_R^0 \rangle = v_4$ and $\langle \Delta_R^0 \rangle = v_6$, 
as already noted, then the minimum of $V_R$ is given by
\begin{equation}
V_0 = m_4^2 v_4^2 + m_6^2 v_6^2 + \frac{1}{2} \lambda_1 v_4^4 + \frac{1}{2} 
\lambda_2 v_6^4 + \frac{1}{2} \lambda_3 v_6^4 + f_1 v_4^2 v_6^2 - 
f_2 v_4^2 v_6^2 + 2 \mu v_4^2 v_6,
\end{equation}
where $v_{4,6}$ are determined by
\begin{eqnarray}
\frac{\partial V_0}{\partial v_4} &=& 2 v_4 [m_4^2 + \lambda_1 v_4^2 + 
(f_1 - f_2) v_6^2 + 2 \mu v_6] = 0, \\
\frac{\partial V_0}{\partial v_6} &=& 2 v_6 [m_6^2 + (\lambda_2 + \lambda_3) 
v_6^2 + (f_1 - f_2) v_4^2] + 2 \mu v_4^2 = 0.
\end{eqnarray}
The physical mass-squared matrices are given by
\begin{eqnarray}
  {\cal M}^2 (Re \phi_R^0, Re \Delta_R^0) &=& \begin{pmatrix} 2 \lambda_1 v_4^2 & 
    2(f_1 - f_2) v_4 v_6 + 2 \mu v_4 \cr 2(f_1 - f_2) v_4 v_6 
    + 2 \mu v_4 & 2 (\lambda_2 + \lambda_3) v_6^2 - \mu v_4^2/v_6\end{pmatrix}, \\ 
  {\cal M}^2 (Im \phi_R^0, Im \Delta_R^0) &=& \begin{pmatrix}-4 \mu v_6 & 2 \mu v_4 \cr
    2 \mu v_4 & -\mu v_4^2/v_6\end{pmatrix}, \\ 
  {\cal M}^2 (\phi_R^\pm, \Delta_R^\pm) &=& \begin{pmatrix}2 v_6 (f_2 v_6 - \mu) & 
    -\sqrt{2} v_4 (f_2 v_6 - \mu) \cr  -\sqrt{2} v_4 (f_2 v_6 - \mu) & 
    v_4^2/v_6 (f_2 v_6 - \mu)\end{pmatrix}, \\ 
  {\cal M}^2 (\Delta_R^{\pm \pm}) &=& 2 f_2 v_4^2 - 2 \lambda_3 v_6^2 
  - \mu v_4^2/v_6.
\end{eqnarray} 
As expected, the linear combinations
$$\frac{(v_4 Im \phi_R^0 + 2 v_6 Im \Delta_R^0)}{\sqrt{v_4^2 + 4 v_6^2}}, ~~~  
\frac{(v_4 \phi_R^\pm + \sqrt{2} v_6 \Delta_R^\pm)}{\sqrt{v_4^2 + 2 v_6^2}}$$
have zero mass, corresponding to the longitudinal 
components of $Z'$ and $W_R^\pm$.  Their orthogonal combinations
$$A_R = \frac{\sqrt{2}(v_4 Im \Delta_R^0 - 2 v_6 Im \phi_R^0)}{ 
  \sqrt{v_4^2 + 4 v_6^2}}, ~~~  \xi_R^\pm = \frac{(v_4 \Delta_R^\pm - \sqrt{2} v_6 
  \phi_R^\pm)}{\sqrt{v_4^2 + 2 v_6^2}}$$
have mass-squares $-\mu(v_4^2 + 4 v_6^2)/v_6$ and 
$(f_2 - \mu/v_6) (v_4^2 + 2 v_6^2)$ respectively.  Since $n_R$ 
couples to $\Delta_R^\pm$, but not to $\phi_R^\pm$, the discussion on 
dark-matter relic abundance from $n n$ annihilation to lepton pairs 
through $\Delta_R^\pm$ exchange in Ref.~\cite{klm09} applies only 
if $v_6^2 << v_4^2$.  This turns out to be exactly what the model 
requires because $m_n$ comes from $v_6$ and $m_n$ of order 200 GeV 
is needed for dark-matter relic abundance.

To be specific, we will assume in fact that $m_n = 200$ GeV.  If this 
value is changed, some details in the following will be changed, but 
all the qualitative features of this model will remain.  The first 
thing to notice is that for $m_n = 200$ GeV, Fig.~3 of Ref.~\cite{klm09} 
requires $m_{\Delta_R^+} = 220$ GeV.  From the Yukawa coupling
\begin{equation}
{f_n \over \sqrt{2}} (\Delta_R^0 n_R n_R + \sqrt{2} \delta_R^+ n_R e_R 
+ \Delta_R^{++} e_r e_R),
\end{equation}
we get $m_n = \sqrt{2} f_n v_6$.  Since $f_n=1$ is assumed in computing 
the relic abundance in Ref.~\cite{klm09}, we obtain $v_6 = 141$ GeV. 
Let us now assume $M_{Z'} = 1$ TeV for illustration.  Then $v_4 = 1851$ GeV 
and $M_{W_R} = 832$ GeV, where $v_2 = 95$ GeV and $v_3 = 146$ GeV have been 
used to ensure zero $Z-Z'$ mixing at tree level (see next section).

\begin{figure}[ht]
  \begin{center}
    \includegraphics[width=15cm]{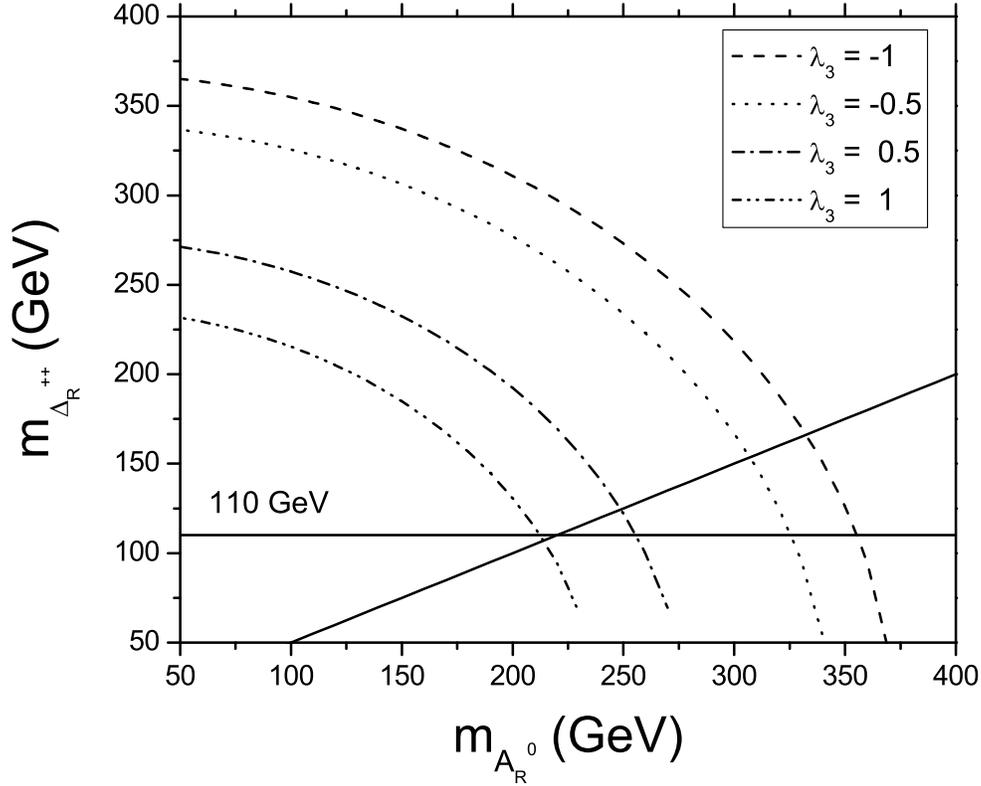}
    \caption{ Plot of $M_{A_R^0}$ versus $M_{\Delta^{++}_R}$ for different 
values of $\lambda_3$ with $v_2 = 95~GeV$, $v_3=146$~GeV, $v_4 = 1851$~GeV, 
$v_6=141$~GeV, $M_{\xi_R^+} =  220$~GeV, and  $M_{Z'} =  1$~TeV, $M_{W_R^+} =  
832$~GeV.  The lower bound of 110 GeV comes from Tevatron data.  The 
line $m_{A_R} = 2 m_{\Delta_R^{++}}$ is also shown.}
    \label{fig:plot-mass}
  \end{center}
\end{figure}

The physical charged scalar $\xi_R^+$ is now 99.4\% $\Delta_R^+$ and its 
mass is given by
\begin{equation}
m^2_{\xi_R^+} = (f_2 - \mu/v_6)(v_4^2 + 2 v_6^2) = [220~{\rm GeV}]^2.
\end{equation}
This implies that $f_2 - \mu/v_6$ = 0.014.  We now note that the 
$\Delta_R$ scalar triplet masses satisfy the important sum rule
\begin{equation}
{m^2_{A_R} \over 1 + 4 v_6^2/v_4^2} + m^2_{\Delta_R^{++}} = 
{2 m^2_{\xi_R^+} \over 1 + 2 v_6^2/v_4^2} - 2 \lambda_3 v_6^2.
\end{equation}
This means that both $m_{A_R}$ and $m_{\Delta_R^{++}}$ are bounded from 
above as a function of $\lambda_3$ which should not be larger than about one 
in magnitude.  We plot in Fig.~\ref{fig:plot-mass} $m_{\Delta_R^{++}}$ versus 
$m_{A_R}$ for various values of $\lambda_3$. 

We now come to a very important conclusion.  To satisfy the dark-matter 
relic density in this model, $m_n$ and $m_{\xi_R^+}$ have to be of order 
200 GeV.  This in turn implies that $m_{A_R}$ and $m_{\Delta_R^{++}}$ are 
bounded in such a way that the decays $A_R \to n n$ and $\Delta_R^{++} \to 
\xi_R^+ \xi_R^+$ are kinematically forbidden.  This means that the dominant 
decay of $\Delta_R^{++}$ is into two like-sign leptons, which is a great 
experimental signature.  There is also an allowed region in parameter 
space which enables the decay $A_R \to \Delta_R^{++} \Delta_R^{--}$. 
Note that the present experimental bound on $m_{\Delta^{++}}$ is 
110 GeV~\cite{tev04}.

As for the remaining two scalar masses from diagonalizing Eq.~(14), 
$H^0_{R2}$ will be heavy with 
$m^2_{H^0_{R2}} = 2 \lambda_1 v_4^2$, whereas $H^0_{R1}$ will be light 
with mass given by
\begin{equation}
m^2_{H^0_{R1}} = {m^2_{A_R} \over 1 + 4v_6^2/v_4^2} + {2 v_6^2 \over \lambda_1} 
[\lambda_1(\lambda_2+\lambda_3) - (f_1 - 0.014)^2].
\end{equation}
Finally, we need to consider the scalar bidoublet and the scalar left 
doublet.  In this model, $(\phi_1^0,\phi_1^-)$ will be heavy, and 
the two doublets $(\phi_2^+,\phi_2^0)$, $(\phi_L^+,\phi_L^0)$ are 
similar to the usual two Higgs doublets considered in the SM. 
The linear combination $(v_3 \Phi_2 - v_2 \Phi_L)/\sqrt{v_2^2+v_3^2} = 
[H_L^+, (H_L^0 + iA_L^0)/\sqrt{2}]$ is physical and light.

\section{Gauge sector}
Since $e$ has $L=1$ and $n$ has
$L=0$, the $W_R^+$ of this model must have $L = S - T_{3R} = 0 - 1 = -1$.
This also means that unlike the conventional LRM, $W_R^\pm$ does not mix
with the $W_L^\pm$ of the SM at all.  This important property allows the 
$SU(2)_R$ breaking scale to be much lower than it would be otherwise, 
as explained already 22 years ago \cite{m87,bhm87}.  Assuming that 
$g_L = g_R$ and let $x \equiv \sin^2 \theta_W$, then the neutral gauge 
bosons of the DLRM (as well as the ALRM) are given by
\begin{equation}
  \begin{pmatrix}A \cr Z \cr Z'\end{pmatrix} = \begin{pmatrix}\sqrt{x} & \sqrt{x} & \sqrt{1-2x} \cr
    \sqrt{1-x} & -x/\sqrt{1-x} & -\sqrt{x(1-2x)/(1-x)} \cr 0 &
    \sqrt{(1-2x)/(1-x)} & -\sqrt{x/(1-x)}\end{pmatrix} \begin{pmatrix}W_L^0 \cr W_R^0 \cr B\end{pmatrix}.
\end{equation}
Whereas $Z$ couples to the current $J_{3L} - x J_{em}$ with coupling
$e/\sqrt{x(1-x)}$ as in the SM, $Z'$ couples to the current
\begin{equation}
  J_{Z'} = x J_{3L} + (1-x) J_{3R} - x J_{em}
\end{equation}
with the coupling $e/\sqrt{x(1-x)(1-2x)}$.  The masses of the gauge bosons
are given by
\begin{eqnarray} \label{bosonmasses1}
&& M_{W_L}^2 = \frac{e^2}{2x} (v_2^2 + v_3^2), ~~~ M_Z^2 = \frac{M_{W_L}^2}{1-x},
~~~ M_{W_R}^2 = \frac{e^2}{2x} (v_2^2 + v_4^2 + 2v_6^2), \\ \label{bosonmasses2}
&& M_{Z'}^2 = \frac{e^2 (1-x)}{2x  (1-2x)} (v_2^2 + v_4^2 + 4v_6^2) -
\frac{x^2 M_{W_L}^2}{(1-x)(1-2x)},
\end{eqnarray}
where zero $Z-Z'$ mixing has been assumed, using the condition
\cite{bhm87} $v_2^2/(v_2^2+v_3^2) = x/(1-x)$.  Note that in the ALRM,
$\Delta_R$ is absent, hence $v_6=0$ in the above.  Also, the assignment
of $(\nu,e)_L$ there is different, hence the $Z'$ of the DLRM is not identical
to that of the ALRM.  At the LHC, if a new $Z'$ exists which couples to
both quarks and leptons, it will be discovered with relative ease.  Once
$M_{Z'}$ is determined, then the DLRM predicts the existence of $W_R^\pm$
with a mass in the range
\begin{equation}
\frac{(1-2x)}{2(1-x)} M_{Z'}^2 + \frac{x}{2(1-x)^2} M_{W_L}^2 < M_{W_R}^2 <
\frac{(1-2x)}{(1-x)} M_{Z'}^2 + \frac{x^2}{(1-x)^2} M_{W_L}^2.
\end{equation}

In the ALRM, since $v_6=0$, $M_{W_R}$ takes the value of the upper limit
of this range.  The prediction of $W_R^\pm$ in addition to $Z'$ distinguishes
these two models from the multitude of other proposals with an extra $U(1)'$
gauge symmetry.

\section{Z' decay}

Consider the possible discovery of $Z'$ at the LHC.  For $M_{Z'} = 1$ TeV, 
only an integrated luminosity of 0.2 fb$^{-1}$ is required~\cite{klm09}. 
Its discovery channel is presumably $\mu^+ \mu^-$, but it will also have 
4 charged muons in the final state from $\Delta_R^{++} \Delta_R^{--}$, 
and perhaps even 8 charged muons, as shown below.

In addition to all SM particles, $Z'$ also decays into $n \bar{n}$, 
$\Delta_R^{++} \Delta_R^{--}$, $\xi_R^+ \xi_R^-$, $A^0_R H_{R1}^0$, 
$H_L^+ H_L^-$, and $A_L^0 H_L^0$.  In particular, the subsequent decay 
$\Delta_R^{\pm \pm} \to \mu^\pm \mu^\pm$ will be a unique signature where 
the like-sign dimuons have identical invariant masses~\footnote{Not all 
models involving doubly charged scalars have this decay, see for 
example~\cite{Aranda:2008ab}.}.

The interactions of $Z'$ with fermions come from
\begin{equation}
{\cal L} = - g' Z'_\mu J^\mu_{Z'},
\end{equation}
where $g' = e/\sqrt{x(1-x)(1-2x)}$.  Ignoring fermion masses, each 
fermionic partial width is given by
\begin{equation}
\Gamma (Z' \to \bar{f} f) = {(g')^2 M_{Z'} \over 24 \pi} [c_L^2 + c_R^2],
\end{equation}
where $c_{L,R}$ are the coefficients from 
$J_{Z'} = x J_{3L} + (1-x) J_{3R} - x J_{em}$, and a color factor of 3 should be  
added for each quark.  In the DLRM, we have
\begin{eqnarray}
&& u_L = -{x \over 6}, ~~~ u_R = {1 \over 2} - {7x \over 6}, ~~~ d_L = -{x 
\over 6}, ~~~ d_R = {x \over 3}, \\ 
&& \nu_L = {x \over 2}, ~~~ n_R = {1-x \over 2}, ~~~ e_L = {x 
\over 2}, ~~~ e_R = -{1 \over 2} + {3x \over 2}.
\end{eqnarray}
Here we need to consider 3 families for $u,d,\nu,e$ but only one for 
$n$. 

The decay of $Z' \to A_R^0 H_{R1}^0$ to scalars come from
\begin{equation}
{\cal L} = - g' (1-x) Z'_\mu [(\partial^\mu H_{R1}^0) A_R^0 - 
(\partial^\mu A_R^0) H_{R1}^0],
\end{equation}
with the partial decay width
\begin{equation}
\Gamma (Z' \to A_R^0 H_{R1}^0) = {(g')^2 M_{Z'} (1-x)^2 \over 48 \pi},
\end{equation}
where $(1-x)$ comes from $I_{3L} = 0, I_{3R}=-1, Q=0$.
For $Z' \to \xi_R^+ \xi_R^-$, the factor is $x$, coming from $I_{3L}=0, 
I_{3R}=0, Q=1$.  For $Z' \to \Delta_R^{++} \Delta_R^{--}$, the factor is 
$(1-3x)$, coming from $I_{3L}=0, I_{3R}=1, Q=2$.

The decay of $Z'$ to the physical Higgs bosons of the effective two-doublet 
electroweak sector should also be considered.  They are $(\phi_2^+,\phi_2^0)$ 
and $(\phi_L^+,\phi_L^0)$.  The physical linear combination is $(v_3 \Phi_2 
- v_2 \Phi_L)/\sqrt{v_2^2+v_3^2}$.  Since $v^2_2/v^2_3 = x/(1-2x)$, the $Z'$ 
couplings are completely determined.  The resulting factor for both 
$Z' \to H_L^+ H_L^-$ and $Z' \to A_L^0 H_L^0$ is $(1-3x)/2$. 

\begin{figure}[ht]
    \includegraphics[width=6cm]{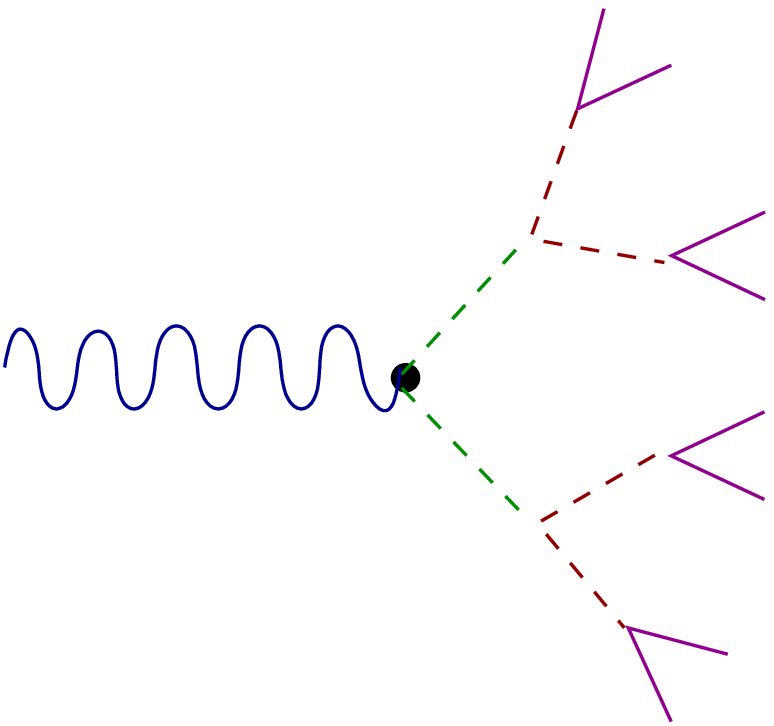}
\caption{ Diagram for the process $Z^{\prime}\to A_R^0 H_R^0 \to 
\Delta_R^{++}\Delta_R^{--} + \Delta_R^{++}\Delta_R^{--} \to 2(\mu^+\mu^+) 
+ 2(\mu^-\mu^-)$}
    \label{fig:zprimedecay}
\end{figure}

Let $\Gamma_0 = (g')^2 M_{Z'}/48 \pi$, then the partial decay widths in 
units of $\Gamma_0$ and their respective branching fractions (\%) are 
given in Table II.
\begin{table}[htb]
\caption{$Z'$ decay widths and branching fractions.}
\begin{center}
\begin{tabular}{|c|c|c|}
\hline
final state & partial width in $\Gamma_0$ & branching fraction (\%)\\
\hline
$\bar{u} u$ & $(9/2)-21x+25x^2 = 0.9925$ & 39.4 \\ 
$\bar{d} d$ & $5x^2/2 = 0.13225$ & 5.3 \\ 
$\bar{\nu} \nu$ & $3x^2/2 = 0.07935$ & 3.2  \\
$\bar{e} e$ & $(1/2)-3x+5x^2 = 0.0745$ & 3.0 \\
$\bar{\mu} \mu$ & $(1/2)-3x+5x^2 = 0.0745$ & 3.0 \\
$\bar{\tau} \tau$ & $(1/2)-3x+5x^2 = 0.0745$ & 3.0 \\
\hline
$\bar{n} n$ & $(1-x)^2/2 = 0.29645$ & 11.8 \\
$A_R^0 H_{R1}^0$ & $(1-x)^2 = 0.5929$ & 23.6 \\
$\xi_R^+ \xi_R^-$ & $x^2 = 0.0529$ & 2.1 \\
$\Delta_R^{++} \Delta_R^{--}$ & $(1-3x)^2 = 0.0961$ & 3.8 \\
$H_L^+ H_L^-$ & $(1-3x)^2/4 = 0.024025$ & 0.9 \\
$A_L^0 H_L^0$ & $(1-3x)^2/4 = 0.024025$ & 0.9 \\
\hline
all & 2.51405 & 100.0 \\
\hline
\end{tabular}
\end{center}
\end{table}
In the special case where $m_{A_R} > 2m_{\Delta_R^{++}}$, which is allowed in 
part of the parameter space shown in Fig.~\ref{fig:plot-mass}, and assuming 
that $m_{H_{R1}^0} > 2m_{\Delta_R^{++}}$ as well, we will have the spectacular 
decay chain  $Z' \to A_R^0 H_{R1}^0 \to \Delta_R^{++}\Delta_R^{--} + 
\Delta_R^{++}\Delta_R^{--}$, resulting in 8 charged muons as shown in 
Fig.~\ref{fig:zprimedecay}.   
This branching fraction is of order 20 percent, given the fact that
both $A_R^0$ and $H_{R1}^0$ decay predominantly into $\Delta_R^{++}\Delta_R^{--}$,
and the dominant decay mode of  $\Delta_R^{\pm \pm}$ is into two charged muons. 
In other parts of the parameter space, the decay $A_R^0 \to \Delta_R^{++} 
\Delta_R^{--}$ is kinematically forbidden, but the branching fraction 
for $Z' \to \Delta_R^{++}\Delta_R^{--}$ is still substantial, yielding 4 muons 
in the final state.

In the above , we have assumed that the $\Delta_R$ scalar triplet couples 
only to muons.  This means that the corresponding scotino $n_\mu$ is part 
of the $SU(2)_R$ doublet $(n_\mu,\mu)_R$ with $m_{n_\mu} = 200$ GeV.  If 
$\Delta_R$ couples to electrons, then $e^+ e^- \to e^+ e^-$ scattering 
through $\Delta_R^{\pm \pm}$ exchange would be much too big to be consistent 
with known data.  We also assume no flavor mixing, i.e. $\Delta_R$ does 
not couple to $\mu e$ for example, or lepton flavor violating processes 
such as $\mu \to eee$ and $\mu \to e \gamma$ would be too big.  However, 
$\Delta_R$ still contributes to the muon anomalous 
magnetic moment which turns out to have the magnitude of the 
experimental discrepancy but of the wrong sign.  To remedy this situation, 
one possibility is to add $SU(2)_L$ fermion doublets $(N,E)_{L,R}$ with $S=0$ 
and a neutral scalar singlet $\chi$ of $S=-1$.  The interaction 
$(\bar{N} \nu_\mu + \bar{E} \mu) \chi$ will contribute positively and 
compensate for $\Delta_R$.  One final complication is that $n_e$ should have 
a mass greater than $n_\mu$ in order that $n_\mu$ is dark matter.  Since it 
cannot come from $\Delta_R$, $n_e$ must have a Dirac mass partner, i.e. 
an $n_L$ singlet.  Of course, we can avoid all constraints by considering 
$n_\tau$ instead as the scotino, in which case $\Delta_R^{++}$ will decay 
into $\tau^+ \tau^+$.  The resulting experimental signature would then 
be much more difficult to pick out.

\section{Conclusion}

We have explored the possible phenomenology of an unconventional $SU(2)_R$ 
model at the TeV scale called the Dark Left-Right Model (DLRM)~\cite{klm09}. 
The scalar sector associated with the $SU(2)_R$ gauge group has been studied 
in detail, including its mass spectrum and its most relevant signature, 
namely, the decay of the doubly charged scalar into same-sign dileptons: 
$\Delta_R^{\pm\pm} \to l^{\pm}l^{\pm}$.  From the requirement of dark-matter 
relic abundance that the $SU(2)_R$ scalar triplet must be relatively light, 
we find that the $Z'$ of this model should decay into them with large 
branching fractions. In particular, $Z' \to \Delta_R^{++} \Delta_R^{--}$ will 
yield 4 charged muons, with 1.3 times the event rate of $Z' \to \mu^+ \mu^-$ 
directly.  More spectacularly, if kinematically allowed, $A_R^0$ and $H_{R1}^0$ 
will decay into $\Delta_R^{++} \Delta_R^{--}$ as well, so that 
$Z' \to A_R^0 H_{R1}^0$ will yield 8 charged muons, with 7.9 times the event 
rate of $Z' \to \mu^+ \mu^-$.  Since a modest luminosity of 0.2 fb$^{-1}$ 
at the LHC will produce 10 dimuon events from this $Z'$ with $M_{Z'}=1$ TeV, 
the predicted events with 4 muons and 8 muons will be clear signals of 
our proposal.

\section*{Acknowledgments}
This work was supported in part by the U.~S.~Department of Energy under
Grant No. DE-FG03-94ER40837, by UC-MEXUS and by CONACYT.

\end{document}